\documentclass[11pt]{article}
\usepackage{acl2012}
\usepackage{times}
\usepackage{latexsym}
\usepackage{amsmath,amsfonts,amssymb}
\usepackage{graphicx}
\usepackage{subfigure}
\usepackage{algorithm}
\usepackage{algorithmic}
\usepackage{multirow}
\usepackage{url}
\usepackage{mdwlist}

\renewcommand{\paragraph}[1]{%
  \smallskip
  \noindent{\bfseries #1}}

\title{A Scale-Space Theory for Text\thanks{1st submitted: Jan. 15 2012; revised: Mar. 28 2012.}}

\author{
  Shuang Hong Yang \\
  College of Computing\\
  Georgia Tech\\
  {\tt shy@gatech.edu}
  }

\date{}

\begin{document}
\maketitle
\begin{abstract}
Scale-space theory has been established primarily by the computer
vision and signal processing communities as a well-founded and
promising framework for multi-scale processing of signals (e.g.,
images). By embedding an original signal into a family of gradually
coarsen signals parameterized with a continuous scale parameter, it
provides a formal framework to capture the structure of a signal at
different scales in a consistent way. In this paper, we present a
scale space theory for text by integrating semantic and spatial
filters, and demonstrate how natural language documents can be
understood, processed and analyzed at multiple resolutions, and how
this scale-space representation can be used to facilitate a variety
of NLP and text analysis tasks.
\end{abstract}

\section{Introduction}
\label{sec:intro}
Physical objects in the world appear differently depending on the
scale of observation/measurement. Take the tree as an example,
meaningful observations range from molecules at the scale of
nanometers, to leaves at centimeters, to branches at meters, and to
forest at kilometers. This inherent property is ubiquitous and holds
equally true for natural language. On the one hand, concepts are
meaningful only at the right resolution, for instance, named
entities usually range from unigram (e.g., ``new") to bigram (e.g.,
``New York"), to multigram (e.g., ``New York Times"), and even to a
whole long sequence (e.g., a song name `` Another Lonely Night In
New York"). On the other hand, our understanding of natural language
depends critically on the scale at which it is examined, for
example, depending on how much detailed we would like to get into a
document, our knowledge could range from a collection of
``\emph{keywords}", to a sentence sketch named ``\emph{title}", to a
paragraph summary named ``\emph{abstract}", to a page long
``\emph{introduction}" and finally to the \emph{entire content}. The
notion of scale is fundamental to the understanding of natural
language, yet it was largely ignored by existing models for text
representation, which include simple bag-of-word (BOW) or unigram
language model (LM), n-gram or higher order LMs, and other more
advanced text/language models \cite{Iyer96,Manning99,Metzler05}. One
key problem with many of these models is their inflexibility ---
they capture the semantic structure rather rigidly at only a single
resolution (e.g., $n$-gram with a single fixed value of $n$).
However, which scale is appropriate for a specific task is usually
unknown a priori and in many cases even not homogeneous (e.g., a
document may contain named entities of different length), making it
impossible to capture the right meanings with a fixed single scale.

Scale space theory is a well-established and promising framework for
multi-resolution representation, developed primarily by the computer
vision and signal processing communities with complimentary
motivations from physics and bio-vision. The key idea is to embed a
signal into the \emph{scale space}, i.e., to represent it as a
family of progressively smoothed signals parameterized by a
continuous variable of \emph{scale}, where fine-resolution detailed
structures are progressively suppressed by the convolution of the
original signal with a smoothing kernel (i.e., a low pass filter
with certain properties) \cite{Wit83,Lin94}.

In this paper, we adapt the scale-space model from image to text
signals, proposing a novel framework that enables multi-resolution
representation for documents. The adaptation poses substantial
challenges as the structure of the semantic domain is nontrivially
complicated than the spatial domains in traditional image scale
space. We show how this can be made possible with a set of
assumptions and simplifications. The scale-space model for text not
only provides new perspectives for how text analysis tasks can be
formulated and addressed, but also enables well-established computer
vision tools to be adapted and applied to text processing, e.g.,
matching, segmentation, description, interests points detection, and
classification. To stimulate further investigation in this promising
direction, we initiate a couple of instantiations to demonstrate how
this model can be used in a variety of NLP and text analysis tasks
to make things easier, better, and most importantly,
scale-invariant.

%%%%   Things not explained due to space limitation   %%%%
% 1. How is it exactly done? What is the overarching idea/principles?
% 2. What are the benefits? Why it works? (flexible variable n-gram with spatial info (long-term and recency dependency))
% 3. How does it differ from pyramid & wavelet? (discrete vs continuous, priorly fixed scale vs infinite scale, computational benefits)
%%%%       Included when more space available         %%%%

\section{Scale Space Representation}
\label{sec:scale-intro}
The notion of scale space is applicable to signals of arbitrary
dimensions. Let us consider the most common case, where it is
applied to 2-dimensional signals such as images. Given an image
$f(x_1,x_2)$, its scale-space representation $\gamma(x_1,x_2,s)$ is
defined by:
\begin{align}
&\gamma(x_1,x_2,s)=f(x_1,x_2)*\ell(x_1,x_2,s)\label{eq1}\\
 =& \int_{\mathbb{R}^2}f(x_1-u_1,x_2-u_2)\ell(u_1,u_2,s)du_1du_2,\nonumber
\end{align}
where $*$ denotes the convolution operator, and
$\ell:\mathbb{R}^2\times\mathbb{R}_+\rightarrow\mathbb{R}$ is a
smoothing kernel (i.e., a low pass filter) with a set of desired
properties (i.e., the scale-space axioms \cite{Lin94}). The
bandwidth parameter $s$ is referred to as scale parameter since as
$s$ increases, the derived image will become gradually smoother
(i.e., blurred) and consequently more and more fine-scale structures
will be suppressed.

It has been shown that the Gaussian kernel is the unique option that
satisfies the conditions for \emph{linear scale space}:
\begin{align}
\vspace{-8pt}
&\ell(\textbf{x},s)=\frac{1}{\sqrt{2\pi
s}}e^{-{(x_1^2+x_2^2)}/{2s}}.
\vspace{-5pt}
\end{align}
The resultant linear scale space representation $\gamma(x_1,x_2,s)$
can be obtained equivalently as a solution to the diffusion (heat)
equation
\begin{align}
\vspace{-4pt}%
&\partial_s\gamma=\frac{1}{2}\triangle \gamma\label{eq3}
\vspace{-4pt}%
\end{align}
with initial condition $\gamma(\textbf{x},0)=f(\textbf{x})$, where
$\triangle$ denotes the Laplace operator which in a 2-dimensional
spatial space corresponds to $\frac{\partial^2}{\partial
x_1^2}+\frac{\partial^2}{\partial x_2^2}$. If we view $\gamma$ as a
heat distribution, the equation essentially describes how it
diffuses from initial value, $f$, in a homogeneous media with
uniform conductivity over time $s$. As we can imagine, the
distribution will gradually approach uniform and consequently the
fine-scale structure of $f$ will be lost.

Scale-space theory provides a formal framework for handling the
multi-scale nature of both the physical world and the human
perception. Since its introduction in 1980s, it has become the
foundation of many computer vision techniques and been widely
applied to a large variety of vision/image processing tasks. In this
paper, we show how this powerful tool can be adapted and applied to
natural language texts.

%We hope to carry these nice properties of scale space from image to
%text.
\section{Scale Space Model for Text}
\subsection{Word-level 2D Image Analogy of Text}
\label{sec:word2d}
A straightforward step towards textual sale space would be to
represent texts in the way as image signal. In this section, we show
how this can be made possible. Other alternative signal formulations
will be discussed in the followed section.

Let $\mathcal{V}=\{v_1,v_2,\ldots,v_M\}$ be our vocabulary
consisting of $M$ words, given a document $d$ comprised of a finite
$N$-word sequence $d=w_1w_2\ldots w_N$, without any information
loss, we can characterize $d$ as a 2D $N\times M$ binary matrix $f$,
with the $(x,y)$-th entry $f(x,y)$ indicates whether or not the
$y$-th vocabulary word $v_y$ is observed at the $x$-th position,
i.e.: $f(x,y)=\delta(w_x,v_y)$, where $\delta(a,b) = 1$ if $a=b$ and
0 otherwise. Hereafter, we will refer to the $x$-axis as
\emph{spatial domain} (i.e., positions in the document,
$x\in\mathcal{X}=\{1,\ldots,N\}$), and $y$-axis as the
\emph{semantic axis} (i.e., indices in the vocabulary,
$y\in\mathcal{Y}=\mathcal{V}$). This representation provides an
image analogy to text, i.e., a document $f$ is equivalent to a
black-and-white image except that here we have \emph{one spatial}
and \emph{one semantic} domains, $(x,y)$, instead of two spatial
domains, $(x_1,x_2)$.

Interestingly, scale-space representation can also be motivated by
this binary model from a slightly different perspective, as a way of
robust density estimation. We have the following definition:

\vspace{5pt} \noindent\textsc{Definition 1.} \emph{A 2D text model}
$f\in\mathbb{R}_+^{N\times M}$ \emph{is a probabilistic distribution
over the joint spatial-semantic space}:
$\mathcal{X}\times\mathcal{Y}\rightarrow\mathbb{R}_+$, $0\leqslant
f(x,y)\leqslant 1$, $\int_x\int_yf(x,y)dxdy=1$. \vspace{5pt}

This 2D text model defines the probability of observing a semantic
word $y$ at a spatial position $x$. The binary matrix representation
(after normalization) can be understood as an estimation of $f$ with
kernel density estimators:
\begin{align}
\vspace{-7pt}%
\textbf{f}(x,\cdot)&=\frac{1}{N}\sum\nolimits_{i=1}^N \delta(w_x-w_i)\textbf{e}_x^\top,\\
\textbf{f}(\cdot,y)&=\frac{1}{M}\sum\nolimits_{j=1}^M\delta(v_y-v_j)\textbf{e}_y,
\vspace{-6pt}%
\end{align}
where $\textbf{e}_i$ is the $i$-th column vector of an identity
matrix, $\textbf{f}(x,\cdot)$ denotes the $x$-th row vector and
$\textbf{f}(\cdot,y)$ the $y$-th column vector. Note that here the
Dirac impulse kernels $\delta$ is used, i.e., words are unrelated
either spatially or semantically. This contradicts the common
knowledge since neighboring words in text are highly correlated both
semantically \cite{Mei08} and spatially \cite{Lebanon07}. For
instance, observing the word ``New" indicates a high likelihood of
seeing the other word ``York" at the next position. As a result, it
motivates more reliable estimate of $f$ by using smooth kernels such
as Gaussian \cite{Wit83,Lin94}, which, as we will see, leads exactly
to the Gaussian filtering used in the linear scale-space theory.
\subsection{Textual Signals}
\label{sec:signal}
The 2D binary matrix described above is not the only option we can
work with in scale space. Generally speaking, any vector, matrix or
even tensor representation of a document can be used as a signal
upon which scale space filtering can be applied. In particular, we
use the following in the current paper:
\begin{itemize*}
\item \emph{Word-level 2D} signal, $f(x,y)$, is the binary matrix we described
in \S\ref{sec:word2d}. It records the spatial position for each
word, and is defined on the joint spatial-semantic domains.
\item\emph{ Bag-of-word 1D} signal is the BOW representation $f(y)=\sum_{x}f(x,y)$, i.e., the 2D matrix
is collapsed to a 1D vector. Since the spatial axis is wiped out,
this signal is defined on the semantic domain alone.
\item \emph{Sentence-level 2D} signal is a compromise between word-level
2D and the BOW signals. Instead of collapsing the spatial dimension
for the whole document, we do it for each sentence. As a result,
this signal, $f(x,y)$, records the position of each sentence; for a
fixed position $x_0$, $f(x=x_0, y)$ records the BOW of the
corresponding sentence.
\item \emph{Topic 1D} signal, $\textbf{f}(x)$, is composed of the
topic embedding of each sentence and defined on the spatial domain
only. Assume we have trained a topic model (e.g., Latent Dirichlet
Allocation) on a universal corpus in advance, this signal is
obtained by applying topic inference to each sentence and recording
the topic embedding {\boldmath{$\theta$}}$_x\in\mathbb{R}^k$, where
$k\ll M$ is the dimensionality of the topic space. Topic embedding
is beneficial since it endows us the ability to address synonyms and
polysemy. Also note that the semantic correlation may have been
eliminated and consequently semantic smoothing is no longer
necessary. In other words, although $\textbf{f}(x)$ is a matrix, we
would rather treat it as a vector-variate 1D signal.
\end{itemize*}
All these textual signals involve either a semantic domain or both
semantic and spatial domains. In the following, we investigate how
scale-space filtering can be applied to these domains respectively.
\subsection{Spatial Filtering}
\label{sec:spatial}
Spatial filtering has long been popularized in signal processing
\cite{Wit83,Lin94}, and was recently explored in NLP by
\cite{Lebanon07,YanZha10}. It can be achieved by convolution of the
signal with a low-pass spatial filter, i.e.,
$\gamma(x,s)=f(x)*\ell(x, s)$. For texts, this amounts to borrowing
the occurrence of words at one position from its neighboring
positions, similar to what was done by a cache-based language model
\cite{JelMerRou91,BeeBerLaf99}.

In order not to introduce spurious information, the filter $\ell$
need to satisfy a set of scale-space axioms \cite{Lin94}. If we view
the positions in a text as a spatial domain, the Gaussian kernel
$\ell(x,s)=\frac{1}{\sqrt{2\pi s}}\exp(-x^2/2s)$ or its discrete
counterpart
\begin{align}
&\ell(n,s)=e^{-s}I_n(s)
\end{align}
are singled out as the unique options that satisfy the set of
axioms\footnote{Including linearity, shift-invariance, semi-group
structure, non-enhancement of local extrema (i.e., monotonicity),
scale-invariance, etc.; see \cite{Lin94} for details and proofs.}
leading to the linear scale space, where $I_n(t)$ denotes the
modified Bessel functions of integer order. Alternatively, if we
view the position $x$ as a time variable as in the Markov language
models, a Poisson kernel $\ell(n,s)=e^{-s}{s^n}/{n!}$ is more
appropriate as it retains temporal causality (i.e., inaccessibility
of future data).

\subsection{Semantic Filtering}
\label{sec:semantic}
Semantic filtering attempts to smooth the probabilities of seeing
words that are semantically correlated. In contrast to the spatial
domain, the semantic domain has some unique properties. The first
thing we notice is that, as semantic coordinates are nothing but
indices to the dictionary, we can permute them without changing the
semantic meaning of the representation. We refer to this property as
\emph{permutation invariance}. Semantic smoothing has been
extensively explored in natural language processing
\cite{Manning99,Zhai04}. Classical smoothing methods, e.g.,
Laplacian and Dirichlet smoother, usually shrink the original
distributions to a predefined reference distribution. Recent
advances explored local smoothing where correlated words are
smoothed according to their interrelations defined by a semantic
network \cite{Mei08}.

Given a semantic graph $\mathcal{G}_v$, where two correlated words
$v_y$ and $v_z$ are connected with weight $\mu_{yz}$, semantic
smoothing can be formulated as solving a graph-based optimization
problem:
\begin{equation}\label{eq7}
\begin{split}
\min_{\gamma}&\quad (1-\lambda) \sum\nolimits_{y=1}^M
\mu_y(\gamma_y-f_y)^2
\\&+ \lambda \sum\nolimits_{y=1}^M\sum\nolimits_{z=1}^M
\mu_{yz} (\gamma_y-\gamma_z)^2,\\
\end{split}
\end{equation}
\noindent where $0\leqslant\lambda\leqslant1$ defines the tradeoff,
$\mu_y$ weights the importance of the node $v_y$. Interestingly, the
solution to Eqn.(\ref{eq7}) is simply the convolution of the
original signal with a specific kernel\footnote{This can be proven
by the first-order optimality of Eq(\ref{eq7}).}, i.e.,
$\gamma=f*\ell$.

Compared with spatial filtering, semantic filtering is, however,
more challenging. In particular, the semantic domain is
heterogeneous and not shift-invariant --- the degree of correlation
$\mu_{yz}$ depends on both coordinates $y$ and $z$ rather than their
difference $(y-z)$. As a result, kernels that provably satisfy
scale-space axioms are no longer feasible. To this end, we simply
set aside these requirements and define kernels in terms of the
dissimilarity $d_{yz}$ between a pair of words $y$ and $z$ rather
than their direct difference $(y-z)$, that is,
$\ell_y(y,z;s)=\ell_x(d_{yz},s)$, where we use $\ell_y$ to denote
semantic kernel to distinguish from spatial kernels $\ell_x$. For
Gaussian, this means $\ell_y(y,z;s)=\frac{1}{\sqrt{2\pi
s}}\exp(-d_{yz}^2/2s)$.
\subsection{Text Scale Space}
\label{sec:textss}
Scale is vital for the understanding of natural language, yet it is
nontrivial to determine which scale is appropriate for a specific
task at hand in advance. As a matter of fact, the best choice
usually varies from task to task and from document to document. Even
within one document, it could be heterogeneous, varying from
paragraph to paragraph and sentence to sentence. For the purpose of
automatic modeling, there is no way to decide \emph{a priori} which
scale fits the best. More importantly, it might be impossible to
capture all the right meanings at a single scale. Therefore, the
only reasonable way is to simultaneously represent the document at
multiple scales, which is exactly the notion of \emph{scale space}.

Scale space representation embeds a textual signal into a
\emph{continuous} scale-space, i.e., by a family of progressively
smoothed signals parameterized by continuous scale parameters. In
particular, for a 2D textual signal $f(x,y)$, we have:
\begin{align}&\gamma(x,y;s_x,s_y)=f(x,y)*\ell(x,y;s_x,s_y),\end{align}
where the 2D smoothing kernel $\ell$ is separable between spatial
and semantic domains, i.e.,
\begin{align}&\ell(x,y;s_x,s_y)=\ell_x(x,s_x)\ell_y(y,s_y).\end{align}
Note that we have two continuous scale parameters, the spatial scale
$s_x\in\mathbb{R}_+$ and the semantic scale $s_y\in\mathbb{R}_+$.
The case for 1D signals are even simpler as they only involve one
type of kernels (spatial or semantic). For a 1D spatial signal
$f(x)$, we have $\ell=\ell_x$, and for a semantic signal $f(y)$,
$\ell=\ell_y$. And if $\textbf{f}$ is a vector-variate signal, we
just apply smoothing to each of its dimensions independently.

\begin{figure}
\centering \mbox{
\subfigure{\includegraphics[width=.23\columnwidth]{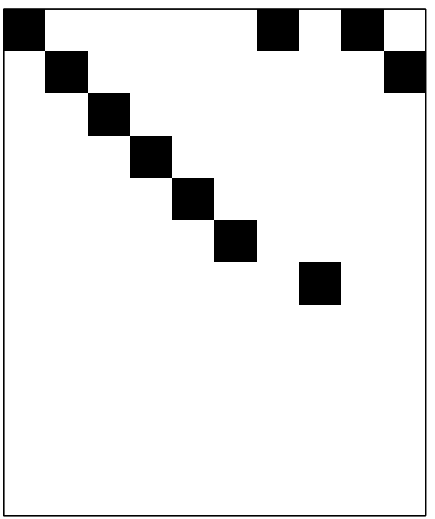}}
\subfigure{\includegraphics[width=.23\columnwidth]{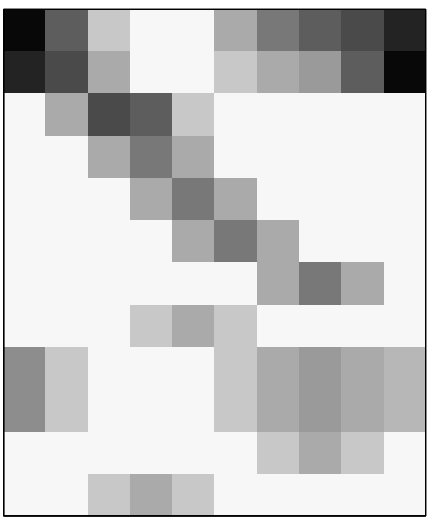}}
\subfigure{\includegraphics[width=.23\columnwidth]{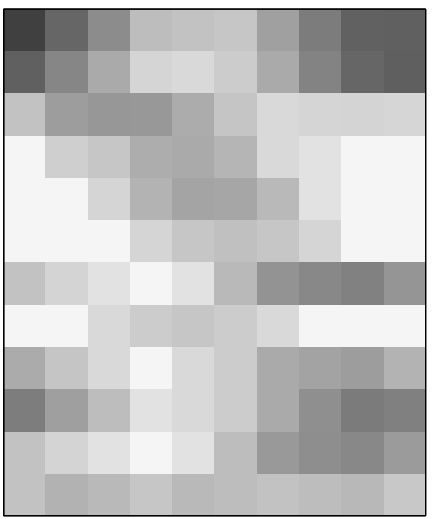}}
\subfigure{\includegraphics[width=.23\columnwidth]{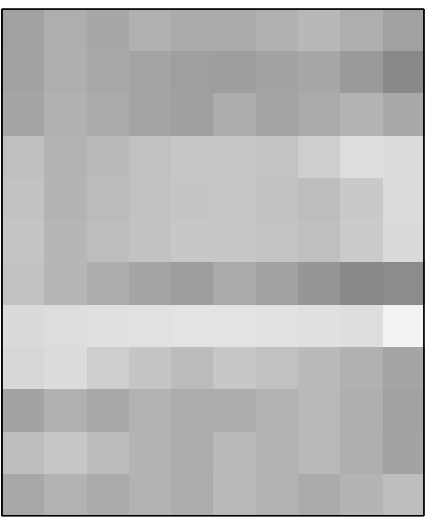}}}
\caption{Samples from the scale space representation of the example
text ``New York Times offers free iPhone 3G as gifts for new
customers in New York" at scales (from left to right): $\textbf{s}$
= $(0,0)$, $(1,1)$, $(4,4)$, and $(64,64)$.\label{fig1}}
\vspace{-10pt}\end{figure}

\paragraph{Example.}
As an example, Figure~\ref{fig1} shows four samples, $\{\gamma(x,y;
\textbf{s}=s_i), ~i=1,2,3,4\}$, from the scale-space representation
$\gamma(x,y;\textbf{s})$ of a synthetic short text ``\emph{New York
Times offers free iPhone 3G as gifts for new customers in New
York}", where $\textbf{s}=(s_x,s_y)$, the two scales are set equal
$s_x=s_y$ for ease of explanation and $\gamma$ is obtained based on
the word-level 2D signal. We use a vocabulary containing 12 words
(in order): ``new", ``york", ``time", ``free", ``iPhone", ``gift",
``customer", ``apple", ``egg", ``city", ``service" and ``coupon",
where the last four words are chosen because of their strong
correlations with those words that appear in this text. The semantic
graph is constructed based on pairwise mutual information scores
estimated on the RCV1-V2 corpus as well as a large set of Web search
queries. The (0,0)-scale sample, or the original signal, is a
$12\times 10$ binary matrix, recording precisely which word appears
at which position. The smoothed signals at (1,1), (2,2) and
(8,8)-scales, on the other hand, capture not only short-range
spatial correlations such as bi-gram, tri-gram and even higher
orders (e.g., the named entities ``New York" and ``New York Times"),
but also long-range semantic dependencies as they progressively
boost the probability of latent but semantically related topics,
e.g., ``iPhone" $\rightarrow$ ``apple", ``customer" $\rightarrow$
``service", ``free" and ``gift" $\rightarrow$ ``coupon", ``new" and
``iPhone" $\rightarrow$ ``egg" (due to the online electronics store
\texttt{newegg.com}).

\section{Scale Space Applications}
\label{sec:experiment}
The scale-space representation creates a new dimension for text
analysis. Besides providing a multi-scale representation that allows
texts to be analyzed in a scale-invariant fashion, it also enables
well-established computer vision tools to be adapted and applied to
analyzing texts. The scale space model can be used in NLP and text
mining in a variety of ways. To stimulate further research in this
direction, we initiate a couple of instantiations.
\subsection{Scale-Invariant Text Classification}
\label{sec:sitk}
In this section, we show how to make text classification
scale-invariant by exploring the notion of \emph{scale-invariant
text kernel} (SITK). Given a pair of documents, $d$ and $d'$, at any
fixed scale $s$, the representation $\gamma$ induces a single-scale
kernel $k_s(d,d^\prime)=\langle \gamma_s, \gamma_{s}^\prime\rangle$,
where $\langle\cdot,\cdot\rangle$ denotes any inner product (e.g.,
Frobenius product, Gaussian RBF similarity, Jensen-Shannon
divergence). This kernel can be made scale-invariant via the
expectation:
{\small%
\begin{align}
&k(d,d^\prime)=\mathbb{E}_q[k_s(d,d^\prime)]=\int_0^\infty
k_s(d,d^\prime)q(s)ds,\label{eq10}
\end{align}
}%
where $q$ is a probabilistic
density over the scale space $\mathbb{R}_+$ with $0\leqslant
q(s)\leqslant1$ and $\int_0^\infty q(s)ds=1$, which in essence
characterizes the distribution of the most appropriate scale. $q$
can be learned from data via a EM procedure or in a Bayesian
framework if our belief about the scale can be encoded into a prior
distribution $q_0(s)$. As an example, we show below one possible
formulation.

Given a training corpus $\mathcal{D}=\{d_i,y_i\}_{i=1}^n$, where $d$
is a document and $y$ its label, our goal in text classification is
to minimize the expected classification error. To simplify matters,
we assume a parametric form for $q$. Particularly, we use the Gamma
distribution $q(s;k,\theta)=\theta^ks^{k-1}e^{-\theta s}/\Gamma(k)$
due to its flexibility. Moreover, we propose a formulation that
eliminates the dependence on the choice of the classifier, which
approximately minimizes the Bayes error rate \cite{Yang08} , i.e.:
\begin{equation}\label{eq11}
\begin{split}
\mathrm{max}_{k,\theta}\; &\sum\nolimits_{i=1}^n\mathbb{E}_q[h_{i}(s)]\\
\end{split}
\end{equation}
where $h_{i}(s)=\triangle_s(d_i,d^m_i)-\triangle_s(d_i,d^h_i)$ is a
heuristic margin; $d^h_i$, called ``nearest-hit", is the nearest
neighbor of $d_i$ with the same class label, whereas $d^m_i$, the
``nearest-miss", is the nearest neighbor of $d_i$ with a different
label, and the distance
$\triangle_s(d,d^\prime)=\sqrt{k_s(d,d)+k_s(d^\prime,d^\prime)-2k_s(d,d^\prime)}$.
This above formulation can be solved via a EM procedure.
Alternatively, we can discretize the scale space (preferably in
log-scale), i.e., $S=\{s_1,\ldots,s_m\}$, and optimize a discrete
distribution $q_j=q(s_j)$ directly from the same formulation. In
particular, if we regularize the $\ell_2$-norm of $\textbf{q}$,
Eq(\ref{eq11}) will become a convex optimization with a close-form
solution that is extremely efficient to obtain:
\begin{equation}\label{eq12}
\textbf{q}=(\bar{\textbf{h}})^+/\vert\vert
(\bar{\textbf{h}})^+\vert\vert
\end{equation}
where $\textbf{q}=[q_1,\ldots,q_m]^\top$, the average margin vector
$\bar{\textbf{h}}=[\bar{{h}}_1,\ldots,\bar{{h}}_m]^\top$ with entry
$\bar{{h}}_j=\frac{1}{n}\sum_{i=1}^nh_{i}(s_j)$, and $(\cdot)^+$
denotes the positive-part operator.

\paragraph{Experiments.}
We test the scale-invariant text kernels (SITK) on the RCV1-v2
corpus with focus on the 161,311 documents from ten leaf-node
topics: \texttt{C11, C24, C42, E211, E512, GJOB, GPRO, M12, M131}
and \texttt{M142}. Each text is stop-worded and stemmed. The top 20K
words with the highest DFs (document frequencies) are selected as
vocabulary; all other words are discarded. The semantic network is
constructed based on pairwise mutual information scores estimated on
the whole RCV1 corpus as well as a large scale repository of web
search queries, and further sparsified with a cut-off threshold. We
implemented the sentence-level 2D, the LDA 1D signals and BOW 1D for
this task. For the first two, the documents are normalized to the
length of the longest one in the corpus via bi-linear interpolation.

We examined the classification performance of the SVM classifiers
that are trained on the \emph{one-vs-all} splits of the training
data, where three types of kernels (i.e., linear (Frobenius), RBF
Gaussian and Jensen-Shannon kernels) were considered. The average
test accuracy (i.e., Micro-averaged F1) scores are reported in
Table~\ref{tab1}. As a reference, the results by BOW representations
with TF or TFIDF attributes are also included. For all the three
kernel options, the scale-space based SITK models significantly
(according to $t$-test at $0.01$ level) outperform the two BOW
baselines, while the sentence level SITK performs substantially the
best with 7.8\% accuracy improvement (i.e., 56\% error reduction).

\begin{table}
\caption{\small Text classification test accuracy. We compared five
models: the bag-of-word vector space models with TF or TFIDF
attributes, and the scale-invariant text kernels with BOW 1D
(SITK.BOW), LDA 1D (SITK.LDA) and Sentence-level 2D (SITK.Sentence)
textual signal. Best results are highlighted in
\textbf{bold}.\label{tab1}}
\begin{center}
{\small
\begin{tabular}{|l|c|c|c|}
\hline Model$\backslash$Kernel  & Linear & RBF & J-S\\
\hline  TF           &0.8789    &0.9087    &0.8901\\
        TFIDF        &0.8821    &0.9099    &0.9016\\
\hline
        SITK.BOW   &0.8917    &0.9143    &0.9076\\
        SITK.LDA   &0.9284    &0.9312    &0.9239\\
        SITK.Sentence   &\textbf{0.9473}    &\textbf{0.9525}    &\textbf{0.9496}\\
\hline
\end{tabular}}
\end{center}
\vspace{-10pt}\end{table}

% TODO: plot scale distribution learned

\subsection{Scale-Invariant Document Retrieval}
\label{sec:silm}
Capturing users' information need from their input queries is
crucially important to information retrieval, yet notoriously
challenging because the information conveyed by a short query is far
more vague and subtle than a BOW model can capture. It is therefore
desirable to base search on more effective text representations than
BOW. We show here how scale space model, together with interest
point detection techniques, can be used to make a retrieval model
scale-invariant and more effective.

Given a set of documents $\{d\}$ and a query $Q$, our goal is to
rank the documents according to their relevance w.r.t. $Q$. The key
to text retrieval is a relevance model $r(Q,d)$. We define $r$ in
the same spirit as we develop the SITK. In particular, if we
normalize the representations of $Q$ and $d$ to the same dimension,
e.g., via bi-linear interpolations\footnote{In the case of the
sentence-level 2D or LDA 1D signals, $\gamma_Q$ is a vector and
$\gamma_d$ is a matrix, this simply amounts to replicating
$\gamma_Q$ to the same dimension as $\gamma_d$, or equivalently
applying a sentence-level sliding-window to $d$, calculating $r$ at
each point and summating the relevance scores.}, then at any fixed
scale $s$, the scale space model induces a relevance function $
r(Q,d\vert s)=\langle\gamma_Q,\gamma_d\rangle$ (e.g., via
KL-divergence, Jessen-Shannon score). This relevance model can be
made scale invariant by defining a distribution $q$ over the scale
space and using:
\begin{align}%
\vspace{-6pt} %
r(Q,d)=\mathbb{E}_q[r(Q,d\vert s)],
\vspace{-6pt}%
\end{align}%
which is referred to as \emph{scale-invariant language model}
(SILM). As in \S\ref{sec:sitk}, $q$ can be learned through a
Bayesian framework or via a EM procedure. As an example, assume $q$
is again a Gamma distribution with parameter $(k,\theta)$. Moreover,
assume we have a training corpus containing a set of queries
$\{Q\}$, and for each $Q$ a set of documents $\{d^Q_j\}$ along with
their relevance judgements. We have the following pairwise
preference learning formulation:
\begin{equation}\label{eq14}
\begin{split}%
\vspace{-6pt} %
&\max_q\; \sum_Q\sum_{d^Q_i\succ d^Q_j}\mathbb{E}_q[h(Q,i,j\vert s)]
\vspace{-6pt} %
\end{split}
\end{equation}
where the pairwise margin $h(Q,i,j\vert s)= r(Q,d^Q_i\vert
s)-r(Q,d^Q_j\vert s)$, and $d^Q_i\succ d^Q_j$ means $d^Q_i$ is more
relevant to $Q$ than $d^Q_j$. This formulation can be solved
efficiently via a similar EM procedure, and again in the discrete
case with $\ell_2$-regularization has an efficient close-form
solution:
\begin{equation}\label{eq15}
\vspace{-1pt} %
\textbf{q}=(\bar{\textbf{h}})^+/\vert\vert
(\bar{\textbf{h}})^+\vert\vert
\vspace{-1pt} %
\end{equation}
where the average margin
$\bar{\textbf{h}}=[\bar{{h}}_1,\ldots,\bar{{h}}_m]^\top$ with
$\bar{{h}}_l=\sum_Q$ $\sum_{d^Q_i\succ d^Q_j}h(Q,i,j\vert s_l)$,
$l=1,\ldots,m$.

More interestingly, scale-space model can also be used, together
with techniques for interest point detection \cite{Lowe04}, to
address passage retrieval (PR) in a scale-invariant manner, i.e., to
determine not only which documents are relevant but also which parts
of them are relevant. PR is particularly advantageous when documents
are substantially longer than queries or when they span a large
variety of topic areas, for example, when retrieving books. A key
challenge in PR is how to effectively narrow our attention to a
small part of a long document. Existing approaches mostly employ a
sliding-window style exhaustive search, i.e., scan through every
possible passage, compute relevance scores and rank all of them
\cite{TelKatLin03}. These approaches suffer from computational
efficiency issues since the number of possible passages could be
quite large for long documents. Here we propose a new idea which
employs interest point detection (IPD) algorithms to quickly focus
our attentions to a small set of potentially relevant passages. In
particularly, for a given ($Q,d$) pair, we first apply IPD (without
normalization) to both $\gamma_Q$ and $\gamma_d$ in scale space,
then match them locally between region pairs centered at each
interest point and calculate the relevance scores there.

\paragraph{Experiments.}
We evaluated SILM on a text retrieval task based on the OHSUMED data
set, a collection of 348,566 documents, 106 queries and 16,140
relevance judgements. Similar preprocessing steps as in
\S\ref{sec:sitk} were implemented. For SILM, standard
Kullback-Leibler divergence was used as relevance function. For
comparison, the unigram language model (i.e., 1-LM) was used as
baselines. The results are reported in Table~\ref{tab2} in terms of
three standard IR evaluation measures, i.e., the
Mean-Average-Precision (MAP), Precision at N with N=5 and 10 (i.e.,
P@5 and P@10). We observe that SILM models outperform the uni-gram
LM amazingly by (up to) 15\% in terms of MAP, 13\% in P@5 and 10\%
in P@10. All these improvements are significant based on a Wilcoxon
test at the level of 0.01. Again, the best performance is obtained
by the sentence-level 2D based SILM model.

\begin{table}
\caption{\small Text retrieval performance. We evaluate four models:
the Unigram Language Model (1-LM) and the Scale-Invariant Language
Models with three textual signal options (referred to as SILM.BOW,
SILM.LDA and SILM.Sentence respectively).\label{tab2}}
\begin{center}
{\small
\begin{tabular}{|l|c|c|c|}
\hline Model$\backslash$Measure  & MAP & P@5 & P@10\\
\hline  1-LM           &0.2699    &0.4812    &0.4659\\
\hline
        SILM.BOW       &0.2807    &0.5076    &0.4762\\
        SILM.LDA       &0.2839    &0.5154    &0.4981\\
        SILM.Sentence  &\textbf{0.3099}    &\textbf{0.5447}    &\textbf{0.5108}\\
        \hline
\end{tabular}}
\end{center}
\vspace{-13pt}\end{table}

\subsection{Hierarchical Document Keywording}
\label{sec:keyword}
The extrema (i.e., maxima and minima) of a signal and its first a
few derivatives contain important information for describing the
structure of the signal, e.g., patches of significance, boundaries,
corners, ridges and blobs in an image. Scale space model provides a
convenient framework to obtain the extrema of a signal at different
scales. In particular, the extrema in the $(k-1)$-th derivative of a
signal is given by the zero-crossing in the $k$-the derivative,
which can be obtained at any scale in the scale space conveniently
via the convolution of the original signal with the derivative of
the Gaussian kernel, i.e.:
\begin{align}
\vspace{-6pt} %
&\frac{\partial^{k}}{\partial x^k}\gamma =
f*\frac{\partial^{k}}{\partial x^k}\ell.
\vspace{-6pt} %
\end{align}
Since Gaussian kernel is infinitely differentiable, the scale-space
model makes it possible to obtain local extrema/derivatives of a
signal to arbitrary orders even when the signal itself is
undifferentiable. Moreover, due to the ``non-enhancement of local
extrema" property, local extrema are created \emph{monotonically} as
we decrease the scale parameter $s$. In this section, we show how
this can be used to detect keywords from a document in a
\emph{hierarchical} fashion. The idea is to work with the word-level
2D signal (other options are also possible) and track the extrema
(i.e., patterns of significance) of the scale-space model $\gamma$
through the zero-crossing of its first derivative $\gamma^\prime=0$
to see how extrema progressively emerge as the scale $s$ goes from
coarse to finer levels. This process reduces the scale-space
representation to a simple ternary tree in the scale space, i.e.,
the so-called ``\emph{interval tree}" in \cite{Wit83}. Since $f$
defines a probability over the spatial-semantic space, it is
straightforward to interpret the identified intervals as keywords.
This algorithm therefore yields a \emph{keyword tree} that defines
topics we could perceive at different levels of granularities from
the document.

\paragraph{Experiments.}
As an illustrative example, we apply the hierarchical keywording
algorithm described above to the current paper. The keywords that
emerged in order are as follows: ``scale space" $\rightarrow$
``kernel", ``signal", ``text"  $\rightarrow$ ``smoothing",
``spatial", ``semantic", ``domains", ``Gaussian", ``filter", ``text
analysis", ``natural language", ``word" $\rightarrow~\ldots$ .

\subsection{Hierarchical Text Segmentation}
\label{sec:seg}
In the previous section, we show how semantic keywords can be
extracted from a text in a hierarchical way by tracking the extrema
of its scale space model $\gamma$. In the same spirit, here we show
how topic boundaries in a text can be identified by tracking the
extrema of the first derivative $\gamma'$.

Text segmentation is an important topic in NLP and has been
extensively investigated previously \cite{BeeBerLaf99}. Many
existing approaches, however, are only able to identify a flat
structure, i.e., all the boundaries are identified at a flat level.
A more challenging task is to automatically identify a
\emph{hierarchical} table-of-content style structure for a text,
that is, to organize boundaries of different text units in a tree
structure according to their topic granularities, e.g., chapter
boundaries at the top-level, followed in order by boundaries of
sections, subsections, paragraphs and sentences as the level of
depth increases. This can be achieved conveniently by the
\emph{interval tree} and \emph{coarse-to-fine tracking} idea
presented in \cite{Wit83}. In particular, if we keep tracking the
extrema of the 1st order derivatives (i.e., rate of changes) by
looking at the points satisfying:
\begin{align}
\vspace{-6pt} %
\frac{\partial^2}{\partial^2 x}\gamma = 0, \text{ while
}\frac{\partial^3}{\partial^3 x}\gamma \ne 0.
\vspace{-6pt} %
\end{align}
Due to the monotonicity nature of scale space representation, such
contours are closed above but open below in the scale space. They
naturally illustrate how topic boundaries appear progressively as
scale $s$ goes finer. And the \emph{exact localization} of a
boundary can be obtained by tracking back to the scale $s=0$. Also
note that this algorithm, unlike many existing ones, does not
require any supervision information.

\begin{figure}
\centering \mbox{
\subfigure{\includegraphics[width=.97\columnwidth,height=2.1cm]{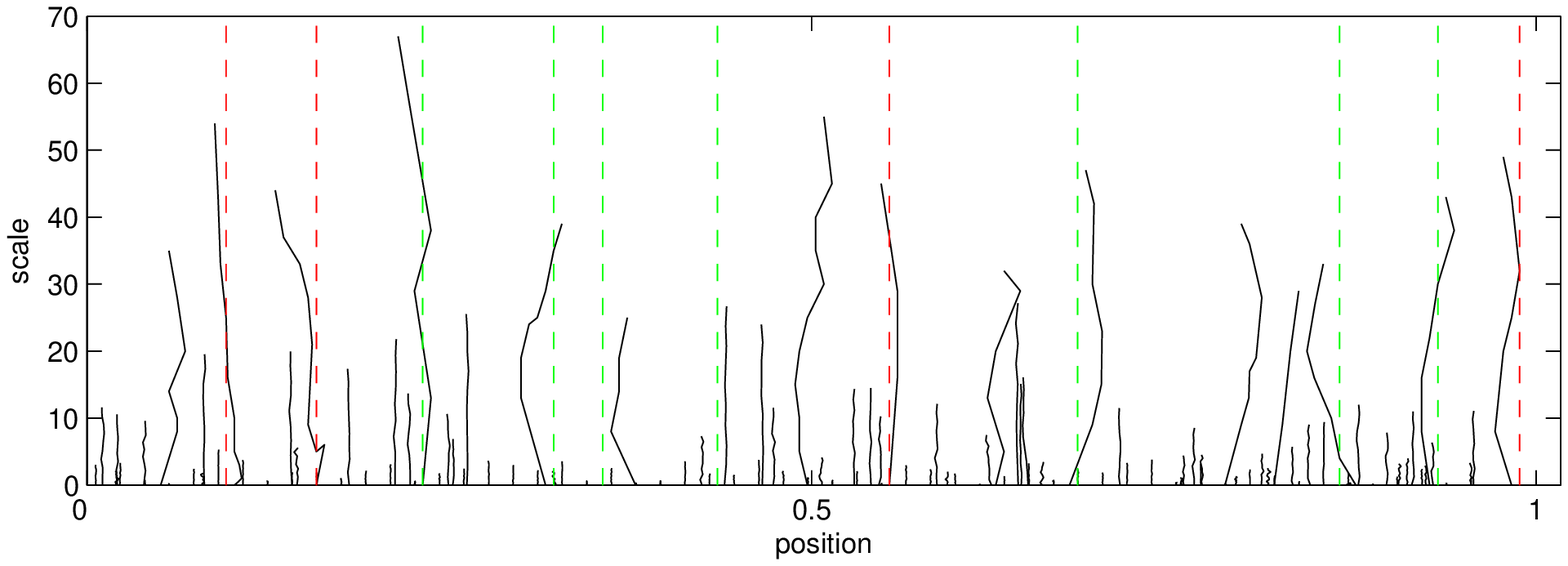}}}
\centering \mbox{
\subfigure{\includegraphics[width=\columnwidth,height=2.1cm]{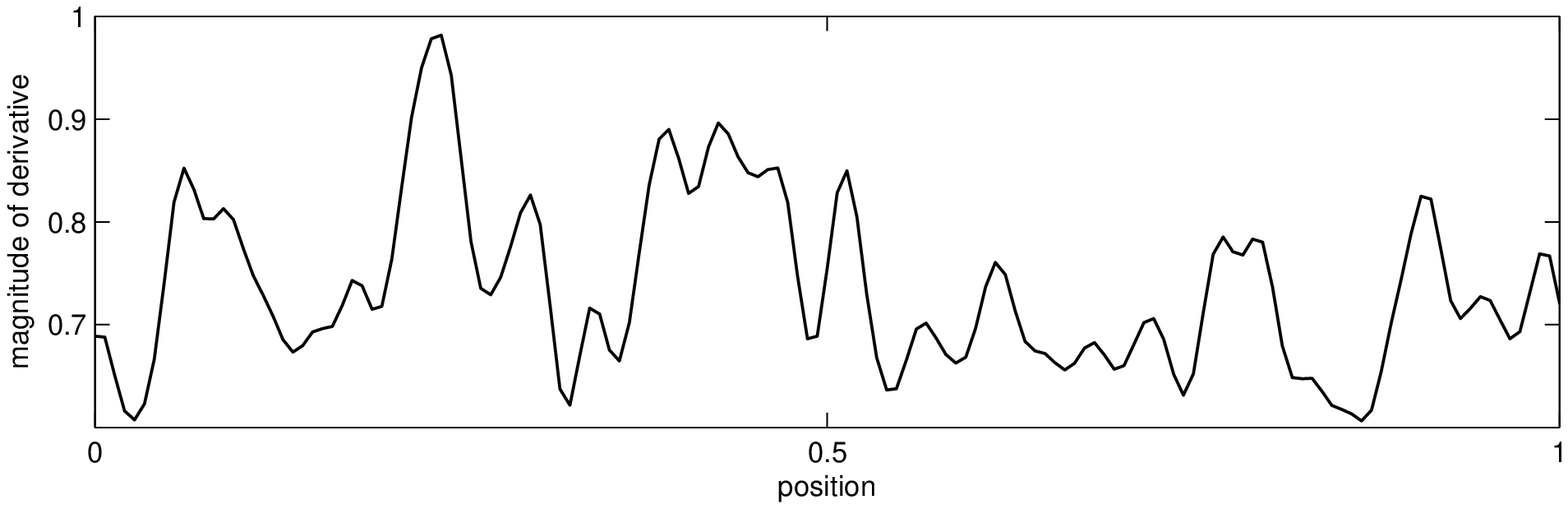}}}
\caption{\vspace{-5pt}Hierarchical text segmentation in scale
space.\label{fig2}} \vspace{-15pt}\end{figure}

\paragraph{Experiments.}
As an example, we apply the hierarchical segmentation algorithm to
the current paper. We use the sentence level 2D signal. Let
{\boldmath{$\gamma$}}$(x,s_x)$ denote the vector
$\gamma(x,\cdot,s_x,s_y=C)$, where the semantic scale $s_y$ is fixed
to a constant $C$, and the semantic index $y$ enumerates through the
whole vocabulary $\{y=v_1,\ldots,v_M\}$. We identify hierarchical
boundaries by tracking the zero contours
$\vert\vert\frac{\partial^2}{\partial
x^2}${\boldmath{$\gamma$}}$(x,s_x)\vert\vert_2=0$ (where
$\vert\vert\cdot\vert\vert_2$ denotes $\ell_2$-norm) to the scale
$s=0$, where the length of the projection in scale space (i.e., the
vertical span) reflects each contour line's topic granularity, as
plotted in Figure~\ref{fig2} (top). As a reference, the velocity
magnitude curve (bottom) $\vert\vert\frac{\partial}{\partial
x}${\boldmath{$\gamma$}}$(x,s_x)\vert\vert_2$, and the true
boundaries of sections (red-dashed vertical lines in top figure) and
subsections (green-dashed) are also plotted. As we can see, the
predictions match the ground truths with satisfactorily high
accuracy.

\vspace{-3pt}
\section{Summary}
\label{sec:sum} \vspace{-2pt}
This paper presented scale-space theory for text, adapting concepts,
formulations and algorithms that were originally developed for
images to address the unique properties of natural language texts.
We also show how scale-space models can be utilized to facilitate a
variety of NLP tasks. There are a lot of promising topics along this
line, for example, algorithms that scale up the scale-space
implementations towards massive corpus, structures of the semantic
networks that enable efficient or even close-form scale-space
kernel/relevance model, and effective scale-invariant descriptors
(e.g., named entities, topics, semantic trends in text) for texts
similar to the SIFT feature for images \cite{Lowe04}.

%\bibliographystyle{acl2012}
%\bibliography{brief}

\begin{thebibliography}{}

\bibitem[\protect\citename{Beeferman \bgroup et al.\egroup }1999]{BeeBerLaf99}
D.~Beeferman, A.~Berger, and J.~Lafferty.
\newblock 1999.
\newblock Statistical models for text segmentation.
\newblock {\em Machine Learning}, 34:177--210.

\bibitem[\protect\citename{Iyer and Ostendorf}1996]{Iyer96}
R.~Iyer and M.~Ostendorf.
\newblock 1996.
\newblock Modeling long distance dependence in language: Topic mixtures vs.
  dynamic cache models.
\newblock {\em IEEE Transactions on Speech and Audio Processing}, 7(1):30--39.

\bibitem[\protect\citename{Jelinek \bgroup et al.\egroup }1991]{JelMerRou91}
F.~Jelinek, B.~Merialdo, S.~Roukos, and M.~Strauss.
\newblock 1991.
\newblock A dynamic language model for speech recognition.
\newblock HLT '1991, pages 293--295.

\bibitem[\protect\citename{Lebanon \bgroup et al.\egroup }2007]{Lebanon07}
G.~Lebanon, Y.~Mao, and J.~Dillon.
\newblock 2007.
\newblock The locally weighted bag of words framework for document
  representation.
\newblock {\em JMLR}, 8:2405--2441.

\bibitem[\protect\citename{Lindeberg}1994]{Lin94}
T.~Lindeberg.
\newblock 1994.
\newblock Scale-space theory: A basic tool for analysing structures at
  different scales.
\newblock {\em Journal of Applied Statistics}, 21(2):224--270.

\bibitem[\protect\citename{Lowe}2004]{Lowe04}
D.~Lowe.
\newblock 2004.
\newblock Distinctive image features from scale-invariant keypoints.
\newblock {\em IJCV}, 60(2):91--110.

\bibitem[\protect\citename{Manning and Schuetze}1999]{Manning99}
C.~Manning and H.~Schuetze.
\newblock 1999.
\newblock {\em Foundations of Statistical Natural Language Processing}.
\newblock MIT Press.

\bibitem[\protect\citename{Mei \bgroup et al.\egroup }2008]{Mei08}
Q.~Mei, D.~Zhang, and C.~Zhai.
\newblock 2008.
\newblock A general optimization framework for smoothing language models on
  graph structures.
\newblock In {\em SIGIR '2008}, pages 611--618.

\bibitem[\protect\citename{Metzler and Croft}2005]{Metzler05}
D.~Metzler and W.~Croft.
\newblock 2005.
\newblock A markov random field model for term dependencies.
\newblock In {\em SIGIR '2005}.

\bibitem[\protect\citename{Tellex \bgroup et al.\egroup }2003]{TelKatLin03}
S.~Tellex, B.~Katz, J.~Lin, A.~Fernandes, and G.~Marton.
\newblock 2003.
\newblock Quantitative evaluation of passage retrieval algorithms for question
  answering.
\newblock In {\em SIGIR '2003}.

\bibitem[\protect\citename{Witkin}1983]{Wit83}
A.~Witkin.
\newblock 1983.
\newblock Scale-space filtering.
\newblock In {\em IJCAI '1983}, pages 1019--1022.

\bibitem[\protect\citename{Yang and Hu}2008]{Yang08}
S.~Yang and B.~Hu.
\newblock 2008.
\newblock Feature selection by nonparametric bayes error minimization.
\newblock In {\em PAKDD '2008}, pages 417--428.

\bibitem[\protect\citename{Yang and Zha}2010]{YanZha10}
S.~Yang and H.~Zha.
\newblock 2010.
\newblock Language pyramid and multi-scale text analysis.
\newblock In {\em CIKM '2010}, pages 639--648.

\bibitem[\protect\citename{Zhai and Lafferty}2004]{Zhai04}
C.~Zhai and J.~Lafferty.
\newblock 2004.
\newblock A study of smoothing methods for language models applied to
  information retrieval.
\newblock {\em ACM TOIS}, 22(2):179--214.

\end{thebibliography}

\end{document}